\documentclass[aps,prb,floats,superscriptaddress,twocolumn,nobibnotes]{revtex4}

\usepackage[dvips]{graphicx}
\usepackage{amsmath}
\usepackage{bm}

\begin{document} 
 
\title{
Metal-insulator transition in EuO
}
\date{\today}  

\author{P.~Sinjukow}
\email{sinjukow@physik.hu-berlin.de}
\author{W.~Nolting}
\affiliation{Lehrstuhl Festk{\"o}rpertheorie, Institut f{\"u}r Physik,
  Humboldt-Universit{\"a}t zu Berlin, Newtonstr.~15, 12489 Berlin}

\begin{abstract}
It is shown 
that 
the spectacular metal-insulator transition in Eu-rich EuO 
can be 
simulated within 
an extended
Kondo lattice model. The different orders of magnitude of the jump in
resistivity in dependence on the concentration of oxygen vacancies as well as
the low-temperature resistance minimum in high-resistivity 
samples 
are reproduced quantitatively.
The huge 
colossal magnetoresistance (CMR)
is calculated and discussed.
\end{abstract}

\maketitle

\section{Introduction}
Stoichiometric (pure) EuO is 
a ferromagnetic semiconductor.
In the sixties and seventies it became famous for
the redshift of its optical absorption edge 
below the Curie temperature $T_C$ (69 K) \cite{Wac79}. 
Eu-rich EuO, whose Eu richness 
is realized by oxygen
vacancies, behaves 
like a metal at low
temperatures 
and shows
a tremendous metal-insulator transition 
near the Curie
temperature \cite{OKDR70,ODMR72,PST72}. A jump in resistivity of 
13 orders of magnitude was
measured by Penney et al.~\cite{PST72} but, since leakage currents limited
the measurement, 
it was 
probably even
greater.
It is a remarkable feature that the size of the jump in
resistivity varies greatly amongst different experimental samples
\cite{ODMR72}.  
Another intriguing 
feature 
is a low-temperature resistance minimum in high-resistivity
samples. 
Furthermore, Eu-rich EuO has a huge colossal magnetoresistance (CMR)
\cite{SFR73}.  

Recently interest
in EuO was renewed by measurements of the resistance and of the
spin-split conduction
band of a EuO film by Steeneken et al.~\cite{STES02}. The authors emphasize the
extraordinary properties of Eu-rich EuO and their importance for basic
research in the field of spintronics due to the
almost perfect spin polarization of the conduction electrons at low
temperatures.

The first 
physical picture (He model) for the metal-insulator transition in
Eu-rich EuO was developed by Oliver et al.\ \cite{OKDR70}. They
postulated a temperature-independent electronic trap (impurity) level
created by the oxygen vacancies. 
It is 
below the
conduction band for high temperatures,
where Eu-rich EuO is an insulator. On decreasing the temperature
below $T_C$, 
the spin-up
conduction band is shifted towards lower energies (redshift) and
therefore crosses 
the impurity level at some point. The electrons of the trap level are
emptied into the conduction band giving rise to EuO 
becoming a metal. 
Later the model was
refined by introducing a Coulomb repulsion between the two electrons 
at an oxygen vacancy and ascribing a
temperature-dependent shift to the trap levels \cite{ODMR72}.
However, the shift of the impurity levels is
unphysical as will be explained below. 
There is a second physical picture (the bound magnetic polaron (BMP)),
which was
introduced by Torrance et al.~\cite{TSM72}. They assumed an exchange
interaction of the surplus electrons of each vacancy site with the Eu
spins. 
Theoretical attempts to quantify the metal-insulator transition 
of Eu-rich EuO were made
by Laks et al. \cite{LaS76}, Mauger \cite{Mau83} and Steeneken
\cite{Ste02}. 
The conductivity in all three cases was
obtained by the simple Drude formula $\sigma=n_c e \mu$ ($n_c$-
conduction-electron density, $e$-elementary charge, $\mu$-mobility). 

In this paper we 
formulate 
an extended Kondo-lattice Hamiltonian as a microscopic model for the 
metal-insulator transition in Eu-rich EuO. It contains qualitatively the
idea of the He model  
through the assumption of impurity trap levels, and it
contains the physics of the BMP 
since an
exchange between the impurity-electron spins and the Eu spins is
included. However, our formulation goes beyond the former two
approaches because of the detailed modelling of the impurity orbital
energies and their hybridization with the conduction band. 
The disorder of the oxygen vacancies is treated within an extension of the
coherent-potential approximation (CPA).
Our calculations 
should be more accurate than the ones in
Refs.~\onlinecite{LaS76,Mau83,Ste02} since we are using a
more complex conductivity formula \cite{MRS92,PaK98}, which 
is derived directly from the fundamental current-current correlation
Kubo formula.

\section{Theory}

In order to model the situation in Eu-rich EuO, first one has to
consider a conduction 
band $\epsilon_{\vec{k}}$. The spins of the 
conduction electrons
$\vec{\sigma}_i$ are coupled to the Eu spins
$\vec{S}_i$ via a coupling constant $J$. These terms represent the
Kondo lattice model. Then, there are impurity levels
$\epsilon_p$, a Coulomb repulsion $U$ 
in case of two electrons at an oxygen vacancy, and an exchange between the
impurity-electron spins $\vec{\sigma}^p_j$ and the Eu spins with coupling
constant $J_p$. Furthermore,
there should be a hybridization $V$ between
impurity electrons ($p_{j\sigma}^\dagger$) and conduction
electrons ($c_{j\sigma}^\dagger$). The superexchange between the Eu
spins is not modeled explicitly.
Instead we are considering $\langle S_z\rangle$ as a parameter, whose
temperature dependence is given by a Brillouin function, which is
fulfilled to very high accuracy in Eu-rich EuO \cite{MGEA78}. 
The Hamiltonian which captures all the features discussed is the
following:
\begin{align}
H=&\sum_{\vec{k}\sigma} \epsilon_{\vec{k}} n_{\vec{k}\sigma} 
- J \sum_i \vec{S}_i \cdot \vec{\sigma}_i 
\nonumber
\\
&+ \sum_{j\sigma} \epsilon_p n^p_{j\sigma}
+ U \sum_j n^p_{j\uparrow} n^p_{j\downarrow} - J_p \sum_{j} \vec{S}_j
\cdot \vec{\sigma}^p_j 
\nonumber
\\
&+ V \sum_{j\sigma} (p_{j\sigma}^\dagger
c_{j\sigma} + c_{j\sigma}^\dagger p_{j\sigma})
\label{eq:Ham}
\end{align}
The sums over $i$ and $j$ mean a sum over all lattice sites and a sum over
the randomly distributed oxygen-vacancy (impurity) sites, respectively.
As the actual value of $J_p$ is not known, $J_p=J$ is assumed for simplicity.

The current operator is given by
\begin{align}
j=\lim_{\vec{q}\to 0} \frac{e}{|\vec{q}|} [H,\rho(\vec{q})]
\end{align}
with the density operator \cite{CzL81}
\begin{align}
\rho(\vec{q})=\sum\limits_{\vec{k}\sigma}
c_{\vec{k}-\vec{q}\sigma}^\dagger c_{\vec{k}\sigma} + 
\sum\limits_{j \sigma}
e^{-i \vec{q} \vec{R}_j}
p_{j\sigma}^\dagger p_{j\sigma} + O (q^2) \;.
\label{eq:dens_op}
\end{align}
Eqs.~(\ref{eq:Ham})-(\ref{eq:dens_op}) yield a surprisingly simple
result for the 
current operator in a cubic system:
\begin{align}
j=-e\sum_{\vec{k}\sigma} \frac{\partial \epsilon_{\vec{k}}}{\partial
  k_x} c_{\vec{k}\sigma}^\dagger c_{\vec{k}\sigma}\;.
\label{eq:curr_op}
\end{align}
Eq.~(\ref{eq:curr_op}) has to be put into the current-current
correlation Kubo formula for the conductivity \cite{Kub57}. If the
self-energy is local, which is exact in infinite
dimensions, e.g.~if 
applying dynamical-mean field theory (DMFT), there is 
a substantial simplification to the Kubo formula. 
All vertex corrections to the current-current correlation function vanish
and one is left with an expression that only contains the one-particle 
spectral density \cite{MRS92,PaK98}. We apply the local approximation to
the self-energy and use the same conductivity formula for the
three-dimensional case:
\begin{align}
\sigma=\frac{e^2 \pi}{\hbar V}\sum_{\vec{k}\sigma}\int\limits_{-\infty}^{\infty} dE \,
(-f^\prime(E)) A_{\vec{k}\sigma}(E)^2
\left(\frac{\partial \epsilon_{\vec{k}}}{\partial k_x}\right)^2  
\label{eq:sigma_DMFT}
\end{align}
where $V$ is the volume, $f^\prime(E)$ is the derivative of the Fermi
function and $A_{\vec{k}\sigma}(E)$ is the spectral density of the conduction
electrons. Eq.~(\ref{eq:sigma_DMFT}) can be
transformed \cite{BHNK01} into 
\begin{align}
&\sigma=\frac{e^2 \pi}{6\hbar a} \sum_\sigma
\int\limits_{-\infty}^{\infty} dE \,
(-f^\prime(E)) \int\limits_{-\infty}^{\infty} dx \,\phi(x,E) \hat{v}(x)
\label{eq:final_cond_formula}
\\
&\phi(x,E)=A_{\vec{k}\sigma}(E)^2_{\epsilon_{\vec{k}}\to x}
\nonumber\\
&\hat{v}(x)=-\int\limits_{-\infty}^{x} dE^\prime\, E^\prime
\rho_0(E^\prime) \;.
\nonumber
\end{align}
$a$ is the lattice constant ($=5.1$ \r{A} in EuO), and 
$\rho_0(E)$ is the free conduction band density of states.

To get the spectral density $A_{\vec{k}\sigma}(E)$, we 
introduce appropriate selfenergies: 
\begin{align}
&\langle\langle [ c_{\vec{k}\sigma},
- J \sum\limits_i \vec{S}_i \cdot
\vec{\sigma}_i ]_-; c_{\vec{k}\sigma}^\dagger \rangle\rangle 
= \Sigma_{\vec{k}\sigma}(E) \, \langle\langle c_{\vec{k}\sigma}; 
c_{\vec{k}\sigma}^\dagger \rangle\rangle 
\label{eq:self-energy_c}
\\ 
&\langle\langle [ p_{j^\prime \sigma},
U \sum\limits_j n^p_{j\uparrow} n^p_{j\downarrow} - J \sum\limits_{j}
\vec{S}_j\cdot \vec{\sigma}^p_j ]_-;
p_{j^\prime \sigma}^\dagger \rangle\rangle   
\nonumber
\\
&\hspace*{3em}
=
\Sigma^p_{\sigma}(E) \, \langle\langle p_{j^\prime \sigma}; p_{j^\prime
  \sigma}^\dagger \rangle\rangle  \;.
\label{eq:self-energy_p}
\end{align}
$\langle\langle\ldots;\ldots\rangle\rangle$ stands for the retarded
Green's function.
Eqs.~(\ref{eq:self-energy_c}) and (\ref{eq:self-energy_p}) are so far
exact relations. 

The 
main approximation of our approach consists in making independent
ansatzes for the conduction-electron and impurity selfenergies. This is
justified if the effect of the hybridization, which to the lowest order
is proportional to $V^2$, is small. In our
calculations it turns out that the hybridization $V$ itself has to be
small 
to give a reasonable order of magnitude of the
metal-insulator transition. Therefore, it is reasonable to neglect
in the selfenergies effects of the order of $V^2$.

The conduction-electron self-energy is taken
from an interpolating ansatz \cite{NRRM01} for the conduction electron
part of the Hamiltonian (\ref{eq:Ham}) ($\sum\limits_{\vec{k}\sigma}
\epsilon_{\vec{k}} 
n_{\vec{k}\sigma}  - J \sum\limits_i \vec{S}_i \cdot \vec{\sigma}_i$):
\begin{align} 
&\Sigma_\sigma(E)=-\frac{1}{2} z_\sigma J \langle S^z \rangle +
  \frac{1}{4} J^2 \frac{a_\sigma G_0(E+\mu-\frac{1}{2} z_\sigma J \langle
    S^z \rangle)}{1-b_\sigma G_0(E+\mu-\frac{1}{2}z_\sigma J \langle
    S^z \rangle)} 
\label{eq:cond_el_self-energy}
\\
&a_\sigma=S(S+1)-z_\sigma \langle S^z \rangle (z_\sigma \langle S^z \rangle +1)\;,\;\; b_\sigma=\frac{1}{2} J
\nonumber
\end{align}
where $z_\sigma=\delta_{\sigma\uparrow}-\delta_{\sigma\downarrow}$ and
$G_0(E)$ is the free conduction electron Green's function.
The ansatz (\ref{eq:cond_el_self-energy}) fulfills all known limiting cases
for $n_c\to 0$ (ferromagnetic saturation, atomic limit, second-order
perturbation theory in $J$ and high-energy-expansion up to the fourth
moment). It is therefore especially
appropriate in the present case of very small conduction-electron
densities ($10^{-3}$ - $10^{-4}$ per unit cell).

For the impurity self-energy $\Sigma^p_{\sigma}(E)$ we take the
self-energy of the  
atomic limit of the correlated Kondo-lattice model \cite{NoM84}
(impurity part of the Hamiltonian (\ref{eq:Ham})) ($\sum\limits_{j\sigma} \epsilon_p
n^p_{j\sigma} + U \sum\limits_j n^p_{j\uparrow} n^p_{j\downarrow} - J_p
\sum\limits_{j} \vec{S}_j \cdot \vec{\sigma}^p_j $). If $U$ is large enough in 
comparison with $J$, this leads to a 4-peak structure in the impurity
quasiparticle density of states with the 
highest peak at
$\epsilon_4=\epsilon_p+U+\frac{J}{2}S$. The peak positions are fixed,
i.e.~do not depend on the magnetization $\langle S_z\rangle$. Therefore,
the above-mentioned picture by Oliver et al.~\cite{ODMR72},
which asigns a 
temperature-dependent shift to the 
impurity levels, is wrong. 
However, the weights of the peaks depend on $\langle
S_z\rangle$ as well as on the impurity occupation. For any parameters
only 3 peaks have a finite weight. The peak at $\epsilon_4$ 
will act as the impurity level which donates electrons
to the conduction band.

We have made ansatzes for the selfenergies
$\Sigma^p_{\sigma}(E)$ and $\Sigma_{\sigma}(E)$ which are correct for
all known limiting cases of $n_c\to 0$ and up to the first order in the
hybridization $V$.

To treat the randomness of the oxygen vacancies we use the
coherent-potential approximation (CPA), which
originally was invented to deal with a non-interacting 
alloy \cite{VKE68}. Here 
we are dealing with strongly interacting
electrons in a two-component alloy. One component corresponds to the
impurity sites $A$, the other to the non-impurity sites $B$. First,
because of our 
effective medium approach and the locality of the impurity self-energy,
local effective one-particle  
energies for the
impurity sites are 
retained. The second problem is how to treat 
the 
non-impurity sites in a consistent way.
To solve this,
for non-impurity sites we also introduce one-particle
levels $\sum\limits_\sigma \epsilon_{p,i} n^p_{i\sigma}$ and hybridization
terms with the conduction band $V \sum\limits_{\sigma} (p_{i\sigma}^\dagger
c_{i\sigma} + c_{i\sigma}^\dagger p_{i\sigma})$. The one-particle
energies $\epsilon_{p,i}$ 
must diverge to infinity at non-impurity sites to assure that these
levels are never really occupied. 
With these 
virtual levels and a non-random
hybridization on the lattice the usual CPA equation
can be formulated: 
\begin{align}
&0=\sum_{m=A,B} c_m
\frac{\epsilon_{p,m}-\mu+\Sigma_{m\sigma}^p(E)-\Sigma_\sigma^{\rm CPA}(E)}{1-G^p_\sigma(E)\left(\epsilon_{p,m}-\mu+\Sigma_{m\sigma}^p(E)-\Sigma_\sigma^{\rm
      CPA}(E)\right) }
\end{align}
\begin{align}
&\epsilon_{p,m}=\left\{ \begin{array}{cc} \epsilon_p & m=A \\ \infty & m=B
  \end{array} \right.\;,\;\;\; \Sigma_{m\sigma}^p=\left\{ \begin{array}{cc}
    \Sigma_\sigma^p  & m=A \\ 0 & m=B
  \end{array} \right.
\nonumber
\end{align}
with $c_A=d$ and $c_B=1-d$ the concentrations of impurity and non-impurity
sites, respectively. $G^p_\sigma$ is the local impurity Green's function
and $\Sigma_\sigma^{\rm CPA}$ the CPA self-energy.
The CPA is known to be correct up to the first order in the impurity
concentration $d$ \cite{Vel69}.
The Green's function of the conduction electrons is given by
\begin{align}
G_{\vec{k}\sigma}(E)=\frac{1}{E-(\epsilon(\vec{k})-\mu)-\frac{V^2}{E-\Sigma_\sigma^{\rm
      CPA}(E)}-\Sigma_\sigma(E)} \;.
\label{eq:Green}
\end{align}
From Eq.~\ref{eq:Green} the spectral density $A_{\vec{k}\sigma}(E)$ can
be obtained. Since $A_{\vec{k}\sigma}(E)^2$ enters the conductivity
formula (\ref{eq:sigma_DMFT}), our calculations of the resisitivity are
correct for all limiting cases of $n_c\to 0$ and up to the second order
in the hybridization $V$ and the impurity concentration $d$.

The parameters for the calculations are as follows.
The width of the conduction band $W$ ($\approx 10{\rm eV}$) can be taken
from the
absorption spectrum of 
Ref.~\onlinecite{STES02} or from band 
structure calculations \cite{Sch00}. We have assumed a semielliptical shape
of the conduction band, which approximates well the actual situation
at the lower band edge \cite{Sch00}. $J$ 
determines the shift of the conduction band ($\frac{J}{2} S$), which 
is about 0.3
eV (half band-splitting) at low temperatures \cite{STES02}. 
For Eu
spins of $S=\frac{7}{2}$ this 
implies a coupling 
constant $J= 0.17{\rm eV}$.
The position of 
$\epsilon_4$ 
is in general not known. Only for the 
highest-resistivity curves an activation energy of 0.3 eV was observed
\cite{ODMR72,TSM72}. If 
$U$ is assumed to be fixed at 1eV, the impurity energy $\epsilon_p$
remains the 
parameter to be adjusted. 
The hybridization $V$, which is also not known, should not
be too large in order not to destroy the effect of the metal-insulator
transition. We have chosen $V=0.01{\rm eV}$. 
Of all parameters, $W$ and $J$ are taken from the experiment. The precise
values of $U$ and $V$ are not important apart from the fact that the
hybridization should be small. The decisive free parameter is the impurity
level $\epsilon_4$ since it determines the position of
the chemical potential. 
For the vacancy (impurity)
concentration $d$ medium values are of the order of 0.1\%
\cite{TSM72,STES02}. 
Two electrons are
assigned to each oxygen vacancy.


\section{Results and Discussion}
\begin{figure}[h]
\includegraphics[width=0.8\linewidth]{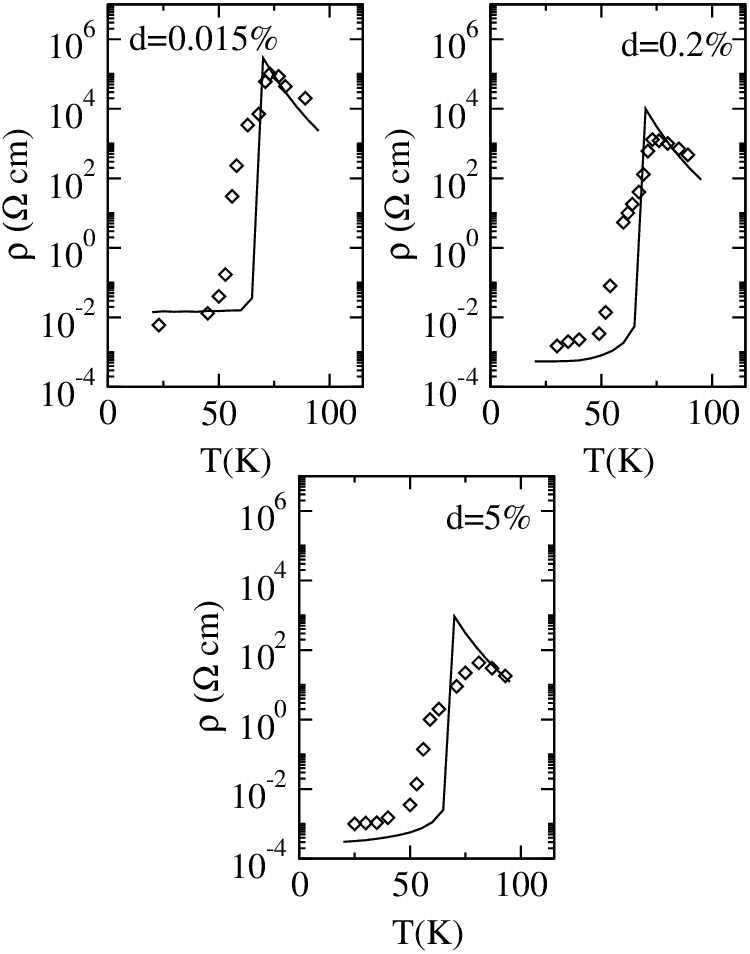}
\caption{Comparison of measured (diamonds, samples 66-6, 95-BA-3 and 49-4 from
  Ref.~\cite{ODMR72}) and 
  calculated resistivity (solid lines) in dependence on temperature. The
  theoretical 
  curves are calculated for three different impurity concentrations
  $d$. $\epsilon_p=-6.44{\rm eV}$ ($\epsilon_4=-5.14{\rm eV}$, lower
  band edge at 
  $-5.30{\rm eV}$ at $T=0$). For other parameters see the text.}
\label{fig:r_T_vgl_exp}
\end{figure}
In Fig.~\ref{fig:r_T_vgl_exp} a comparison 
of
calculated and 
measured resistivity curves is shown. The measured curves represent
three examples of moderate resistivity from
Ref.~\onlinecite{ODMR72}. 
The only varying parameter of the three theoretical curves is the
vacancy concentration $d$.   
$\epsilon_p$ (and with it $\epsilon_4$)
was fixed so as to yield the 
best overall agreement 
with the experimental curves. 
A variation of $\epsilon_p$ ($\epsilon_4$) has
to be taken into 
account in order to explain the resistance minimum in high-resistivity
samples (see below). 
Each curve in Fig.~\ref{fig:r_T_vgl_exp} shows a huge
metal-insulator transition of several orders of magnitude near the
magnetic transition temperature. 
The resistivity in the whole temperature range gets higher and the
jump in resistivity at $T_c$ bigger, the fewer
oxygen vacancies (electron donors) there are. 
Note the different orders of magnitude (7, 6 and
5) of the jump in resistivity of the three experimental
samples, which together with the absolute values of the resistivity 
are reproduced fairly well by the calculated curves. 
This demonstrates that the variation in
the resistivity behaviour can be explained by different
impurity concentrations. 
Incidentally, 
there is almost a $100 \%$ spin
polarization of the conduction electrons at low temperatures as reported
by Steeneken et al.\ \cite{STES02}, which is important for
possible applications in spintronics.

\begin{figure}[h]
\includegraphics[width=\linewidth]{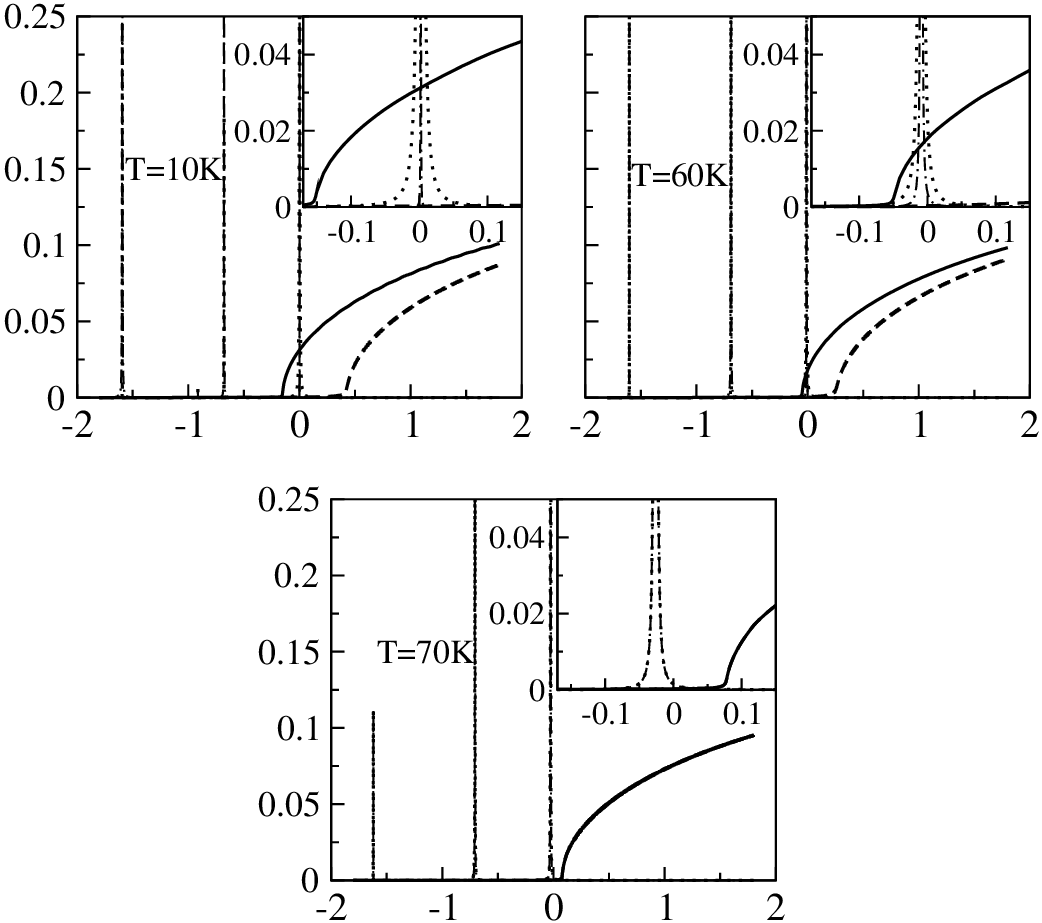}
\caption{Conduction-electron densities of states ($\rho_\uparrow(\mbox{---}),
  \rho_\downarrow(--)$) and impurity densities of states
  ($\rho_{p\uparrow}(-\cdot-), \rho_{p\downarrow}(\cdots)$) for
  $\epsilon_p=-6.44 {\rm eV}$, $d=0.05\%$ and three different
  temperatures. For other parameters see the text. The insets show the densities
  of states near the Fermi level.} 
\label{fig:qdosI_2}
\end{figure}

A {\it qualitative} explanation of the metal-insulator transition is possible in
terms of quasiparticle densities of states along the lines of the He
model (with the difference that there are three instead of one impurity
level, so the number of electrons which can be emptied into the
conduction band is not so big). As shown in
Fig.~\ref{fig:qdosI_2} for low temperatures the uppermost impurity level
$\epsilon_4$ lies within the spin-up part of the conduction band, which
is down-shifted due to the exchange with the Eu spins. Therefore, a part
of the impurity electrons is emptied into the conduction band. The
system is a metal. On increasing the temperature the conduction band is
shifted upwards. The number of conduction electrons decreases and the
resistivity rises. Above the Curie temperature the conduction band is
well separated from the uppermost impurity level, so only thermally
excited electrons contribute to the conductivity, like in an n-doped
semiconductor. The resistivity is
high and decreases with increasing temperature. Hence, the system is
an insulator. 

For a {\it quantitative} analysis of the metal-insulator
transition we rely on the
conductivity formula Eq.~(\ref{eq:final_cond_formula}). 
The energy integrals in that formula yield a highly non-linear
dependence of the conductivity 
on the number of 
conduction electrons $n_c$, which goes beyond the simple Drude formula
$\sigma=n_c e \mu$ applied by other authors.

For high-resistivity samples there is a characteristic
low-temperature minimum in the resistivity \cite{ODMR72}. 
\begin{figure}[h]
\includegraphics[width=0.7\linewidth]{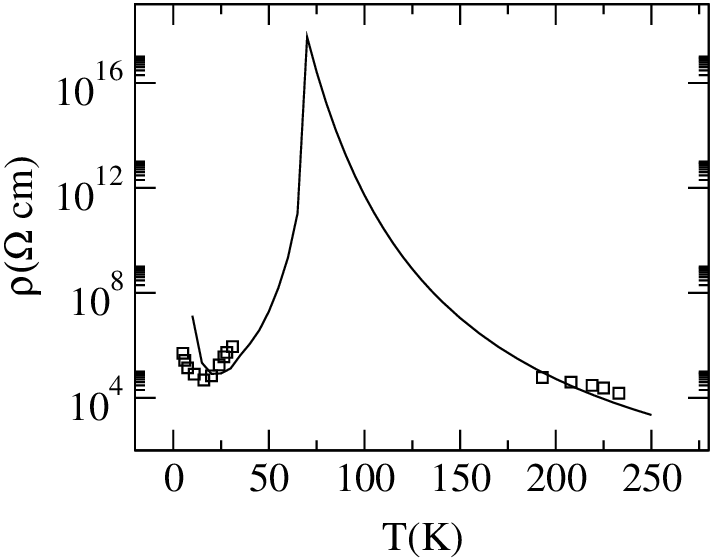}
\caption{Comparison of measured (squares, sample 34-2-30 from
  Ref.~\cite{ODMR72}) and calculated 
  resistivity (solid line) for a high-resistivity sample. Impurity
  concentration 
  $d=0.01\%$, $\epsilon_p=-6.61 {\rm eV}$ ($\epsilon_4=-5.31 {\rm eV}$,
  lower band edge at $-5.30{\rm eV}$ at $T=0$). For other parameters see the text.}
\label{fig:r_T_min}
\end{figure}
Oliver et al.~\cite{ODMR72} explained it by activation from
additional impurities like Gd or 
La. However, a more direct explanation is possible within our model
without assuming additional impurities. The
high values of the resistivity indicate that the impurity peak at 
$\epsilon_4$ must be rather distant from the conduction band at $T_C$. On
lowering the temperature, the resistivity first decreases since the
spin-up conduction band moves down towards the impurity level
$\epsilon_4$. More electrons are thermally excited from the impurity
level into the conduction band. 
If the position of $\epsilon_4$
is low enough never to cross the conduction band, then on lowering the
temperature further, the resistivity increases again near the 
ferromagnetic saturation $\langle S_z\rangle \approx S$. When the
conduction band stays almost fixed with respect to the impurity level,
fewer electrons are excited into the conduction band with decreasing
temperature. 
To test this explanation, we have calculated the temperature-dependent
resistivity for $\epsilon_p=-6.61{\rm eV}$ ($\epsilon_4=-5.31{\rm eV}$,
lower band edge at $-5.30{\rm eV}$ at $T=0$) 
and impurity concentration $d=0.01\%$.
In Fig.~\ref{fig:r_T_min} the calculated and a measured \cite{ODMR72}
curve for a 
high-resistivity sample are shown.
The theoretical curve has a minimum at $T=20{\rm{}K}$ which fits quite
well to the experimental one. Moreover, 
there is a quantitative agreement between the high-temperature tails of
the experimental and the calculated curves. For the temperature range in
between no measured points 
are available but the
run of the theoretical curve seems credible although it takes on very
high values. (Penney et al.'s above-mentioned resistivity measurements
\cite{PST72} were limited to values 
of $10^{11} \Omega {\rm cm}$ but interpolating their
data allows values of up to $10^{16} \Omega {\rm cm}$.)

\begin{figure}[h]
\includegraphics[width=0.8\linewidth]{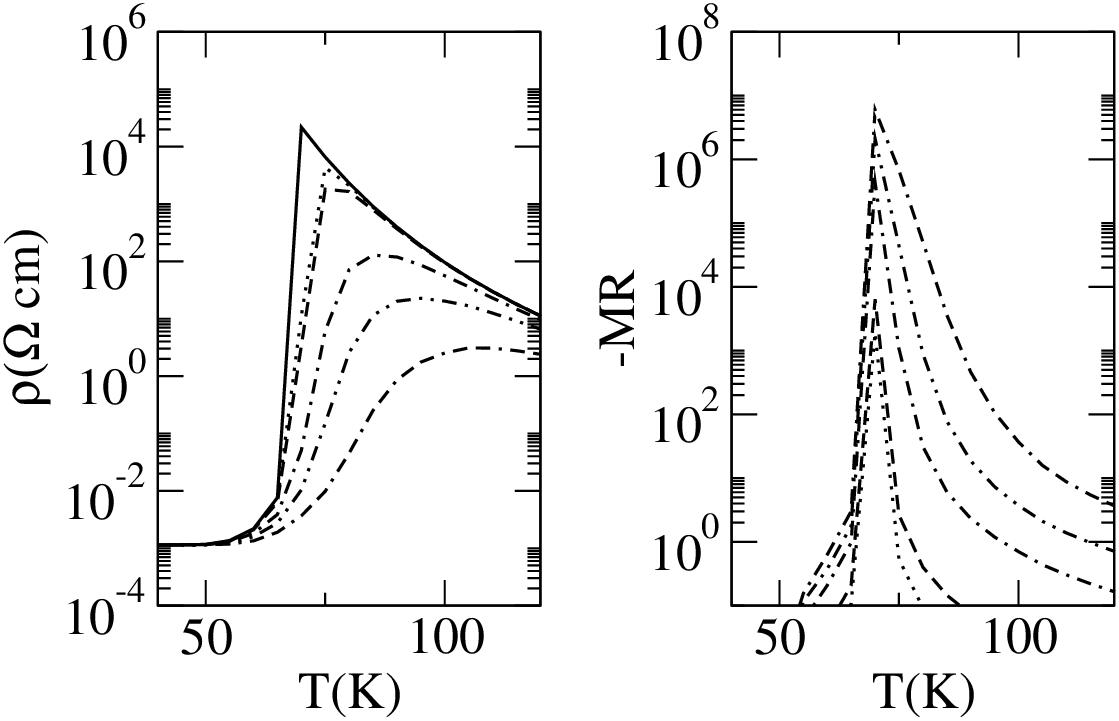}
\caption{Left: Calculated resistivity for
  different magnetic fields $B$; right: negative magnetoresistance. $B=$
  0T(---), 
0.05T ($\cdots$), 0.1T ($- -$), 0.5T (-$\cdot$-$\cdot$), 1T (-\!
$\cdot\cdot$ \!-\!\,$\cdot\cdot$), 2T ($\cdot -\!- \cdot$). $d=0.1\%$;
for other
parameters see the text.} 
\label{fig:r_T_B}
\end{figure}

The calculated dependence of the resistivity on a magnetic
field $B$ is shown 
in the left graph of Fig.~\ref{fig:r_T_B}. It looks qualitatively
similar to the measured dependence in Fig.\ 3 of Ref.~\onlinecite{SFR73}. A
characteristic shift of the 
resistivity maximum from $T_C$ towards higher temperatures can be
observed. This is due to the shift of the conduction band 
depending on the magnetization. Applying a magnetic field has a similar
effect as lowering the temperature. At the  
same time the value of the maximum decreases significantly since 
the metal-insulator transition is smeared out because of the
higher temperature of the maximum. Therefore
there is a huge negative magnetoresistance ${\rm
  MR}=\frac{\rho(B)-\rho(0)}{\rho(B)}$ as shown in the right graph of
Fig.~\ref{fig:r_T_B}. 
If one normalizes the magnetoresistance with $\rho(0)$
instead of $\rho(B)$, the magnetoresistance will
be practically 1 in a wide temperature range. This 
can be compared
with a value of only 0.8 for the colossal
magnetoresistance (CMR) of the manganites like ${\rm
  La}_{1-x}{\rm Ca}_x{\rm Mn O}_3$ \cite{CVM99}. 

\section{Conclusions}
In summary, a model has been presented which, in connection with a
conductivity formula that is derived from the fundamental Kubo formula
in the local-self-energy approximation, allows to accurately
reproduce the spectacular metal-insulator transition and the huge CMR in
Eu-rich EuO. The oxygen vacancy concentration is the decisive parameter for 
the big variations in resistivity behaviour. It may influence the
position of the uppermost impurity level, which is important in order to
explain the resistance minimum in high-resistivity samples. The precise
dependence of the resistivity on both the impurity concentration and the
position of the impurity level should be a subject of thorough future
experimental investigation, especially if aiming at applications in the
field of spintronics.

\begin{acknowledgements}
This work was supported by the Deutsche Forschungsgemeinschaft (Sfb 290).
\end{acknowledgements}

\end{document}